\documentclass[twocolumn,prb,aps,showpacs,groupedaddress]{revtex4-1}
\usepackage{graphicx}
\usepackage{amssymb}
\usepackage{textcomp}
\usepackage{amsmath}
\usepackage[normalem]{ulem} 
\usepackage{soul} 
\def\be{\begin{equation}}
\def\ee{\end{equation}}
\def\bea{\begin{eqnarray}}
\def\eea{\end{eqnarray}}
\def\bdm{\begin{displaymath}}
\def\edm{\end{displaymath}}
\def\ba{\begin{array}}
\def\ea{\end{array}}

\begin{document}




\title{Manifestations of topological band crossings in bulk entanglement spectrum: An analytical study for integer quantum Hall states}

\author{Chi-Ken Lu}
\email{Lu49@ntnu.edu.tw}
\author{Dah-Wei Chiou}
\email{dwchiou@gmail.com}
\author{Feng-Li Lin}
\email{linfengli@ntnu.edu.tw}
\address{Physics Department, National Taiwan Normal University, Taipei 11677, Taiwan}

\date{\today}

\begin{abstract}

We consider integer quantum Hall states and calculate bulk entanglement spectrum by formulating the correlation matrix in guiding center representation. Our analytical approach is based on the projection operator with redefining the inner product of states in Hilbert space to take care of the restriction imposed by the (rectangle-tiled) checkerboard partition. The resultant correlation matrix contains the coupling constants between states of different guiding centers parameterized by magnetic length and the period of partition. We find various  band-crossings by tuning the flux $\Phi$ threading each chekerborad pixel and by changing filling factor $\nu$. When $\nu=1$ and $\Phi=2\pi$, or $\nu=2$ and $\Phi=\pi$, one Dirac band crossing is found. For $\nu=1$ and $\Phi=\pi$, the band crossings are in the form of nodal line, enclosing the Brillouin zone. As for $\nu=2$ and $\Phi=2\pi$, the doubled Dirac point, or the quadratic point, is seen. Besides, we infer that the quadratic point is protected by C$_4$ symmetry of the checkerboard partition since it evolves into two separate Dirac points when the symmetry is lowered to C$_2$.  In addition, we also identify the emerging symmetries responsible for the symmetric bulk entanglement spectra, which are absent in the underlying quantum Hall states.







\end{abstract}


\maketitle

\section{Introduction}

  Entanglement spectrum, which provides finer characteristics for the many-body entanglement than the entanglement entropy, has been used to characterize the topological quantum states.~\cite{Li} One characteristic for these states is the existence of the gapless mode in the entanglement spectrum.~\cite{Grover,Spain1,Turner,Fidkowski,Arovas} This reflects the entangled nature of the underlying quantum state as there is no dominant term in the wavefunction to be approximated as a product state.  On the other hand, gapless excitations at edges are the main character distinguishing the gapped systems that are topological from those that are not. The Dirac band crossings found on the surface of topological insulators~\cite{Haldane,exp} are examples of electron systems.  The partition boundary for evaluating the entanglement spectrum, albeit artificial, plays the similar role as the physical edge hosting the topological gapless edge modes. Then, the correlation between entanglement and the topological order motivated the work~\cite{Qi-Ludwig} to demonstrate a general relation between entanglement spectrum and the physical edge state spectrum of topological states. 

   The entanglement spectrum, by construction, depends on how we partition the ground state in either real space or internal space. The simplest partition is just to cut the system into two halves by a single boundary in a real space, which is also adopted in most of the studies mentioned above. This partition usually breaks some of the underlying translational or discrete symmetries, especially for the systems living on the lattice. It is then interesting to see what the new feature of the entanglement Hamiltonian and the relation with the physical edge Hamiltonian of the topological state will arise if we partition in a different way to preserve some lattice structure.  Because of the residual lattice structure left by the partitioning, one naturally expects the emergence of the band-like structures for the entanglement spectrum, and the ensured gapless modes for the entangled topological state lead to the expectation of the band-crossings.  
   
   Along the line, the work~\cite{MIT1} studied the integer quantum Hall ground state on a square lattice and its entanglement spectrum under a checkerboard partition. By doing so the translational symmetry is reduced to that of a superlattice, and their remarkable findings include a family of entanglement spectra which shows how the gaps are evolved to zero at a point when the equal and extensive partition is employed. Under such extensive partition, the {\it{bulk entanglement spectrum}} (BES) is thus coined. The subsequent work~\cite{Wan1} has also confirmed the the existence of a Dirac band crossing at $\Gamma$ point when total Chern number of the underlying ground state is one. The debate on whether a gap is open or a quadratic band crossing emerges when the total Chern number is even can be clarified if further analytical investigations are taken, which is one purpose of the present paper. More recently, implementation of extensive partition to one-dimensional interacting system has also been made.~\cite{Wan2,MIT2,MIT3,Wan3} 
   
     The topological insulators/superconductors are typically characterized by the patterns of the Dirac band crossings at the physical edges, the robustness of which is protected by some discrete symmetries such as particle-hole, time-reversal and chiral symmetries.~\cite{Furusaki,Kitaev} Based on the similarity between the entanglement spectrum and physical edge spectrum for the topological states, one may expect that the band-crossings of BES could be also protected by some discrete symmetries. It is then interesting to find out the explicit form of entanglement spectrum for some underlying topological quantum state and then identify the emerging discrete symmetries. 
 
    In this paper, we address the above issues by considering the integer quantum Hall states under a checkerboard partition in continuous space, which allows an analytical investigation and gains more insights than previous numerical works do. Given the exact relation between correlation matrix for free-electron system and entanglement spectrum,~\cite{Chung1,Peschel,Cheong} it suffices to focus on the correlation matrix. Hence, we introduce the notion of guiding center representation in which the inner product of states is redefined to take care of the underlying partition. The resultant correlation matrix resembles the one-dimensional hopping Hamiltonian. Besides, the matrix is always in the form of an identity matrix times $\frac{1}{2}$ plus a purely imaginary matrix, from which we immediately conclude that the complex conjugation is the particle-hole symmetry of BES. We further demonstrate the BES for $\nu=1$ and $\nu=2$ cases and find the Dirac and quadratic band crossings when $\Phi=2\pi$, which is similar to the spectra in single layer graphene~\cite{graphene} and bilayer graphene.~\cite{McCann}  In addition, the quadratic point is inferred to be protected by C$_4$ symmetry since it is evolved into two separate Dirac points as we lower the symmetry of partition into C$_2$. The breaking of one quadratic point into two Dirac points also occurs in the nematic ordered states in bilayer graphene.~\cite{nematic} Moreover, the lone Dirac point can also be found in $\nu=2$ and $\Phi=\pi$, while the band crossing become the nodal line in the case of $\nu=1$ and $\Phi=\pi$, which is reminiscent of the nodal structure in Weyl semimetal.~\cite{weyl}

The paper has the following organizations. In Sec.~\ref{formulation} the notion of guiding center representation for correlation matrix is introduced with a simpler example in left-right partition. The formulation to obtain the correlation matrix in checkerboard partition then follows. In Sec.~\ref{espectrum}, the flux through  each square is taken as $2\pi$ and we demonstrate the BES of $\nu=1$ and $\nu=2$ integer quantum Hall states where the band crossing points are found. The associated discrete symmetries of the entanglement Hamiltonian are also identified. We also consider the case of half-flux and obtain the nodal lines in BES. In the end, the conclusions is drawn in Sec.~\ref{discussion}.

\section{Correlation matrix in guiding center representation}\label{formulation}

 To obtain the entanglement spectrum for a underlying quantum state $|\Psi\rangle$, we should first construct the reduced density matrix $\rho_A$ for the subsystem $A$. It is given by $\rho_A:=\textrm{Tr}_B |\Psi\rangle\langle \Psi|$ where we have traced out the Hilbert space of the subsystem $B$ if we partition the total system into subsystems $A$ and $B$. Then the entanglement Hamiltonian is $H_e:=-\ln \rho_A$ whose eigenvalues form the entanglement spectrum. 
 
  The entanglement Hamiltonian is usually nonlocal and hard to obtain in closed form even for gaussian systems. However, in the case of free fermion system, the usual approach to obtain the entanglement spectrum, the set of eigenvalues of entanglement Hamiltonian $H_e$, is through the correlation matrix $C$ with the aid of the following equality~\cite{Cheong,Peschel}
\be
	H^T_e = \ln\frac{1-C}{C}\:,\label{Cheong1}
\ee in which $H^T_e$ denotes the transpose of $H_e$. The information of partition of system is clearly seen in the coordinate representation for correlation matrix $C(\bf r,r')$ with $\bf r$ and $\bf r'$ both lie within the focused subsystem, say $A$. In fact $C(\bf r,r')$ is the matrix element of the projection operator onto the occupied states,~\cite{MIT1,Fukui,Fukui1}

\be
	P=\sum_{E_a<E_f} |a\rangle \langle a|\:,\label{projection1}
\ee in which the eigenstate $|a\rangle$ of physical Hamiltonian is indexed by the quantum number $a$. In a lattice, $\bf r$ and $\bf r'$ shall run through a subset of sites, so the correlation matrix $C$ is a finite one. In continuous space, one may write down $C$ in coordinate representation. The eigenvalue problem of $C$ then becomes an integral equation. A simple example is available from the integer quantum Hall state~\cite{Spain1,Turner} with left-right partition. It was shown~\cite{Spain1} that for filling factor $\nu=1$ the correlation matrix in coordinate representation is $C({\bf r,r'})=\exp{[-(y-y')^2/4-(x-x')^2/4-i(x+x')(y-y')/2]/2\pi}$, and the following integral equation can be solved analytically,

\be
	\int_{{\bf r,r'}\in A} d{\bf r'}C({\bf r,r'}) f({\bf r'}) = \alpha f({\bf r})\:,\label{integral_eq}
\ee with $A$ denote the subsystem. In fact, the eigenfunction $f$ of Eq.~\ref{integral_eq} coincides with the lowest Landau level wavefunction, and the corresponding eigenvalue $\alpha$ is equal to the probability of locating $f$ in the subsystem $A$. 

\begin{figure}
\input{epsf}
\includegraphics[width=0.3\textwidth]{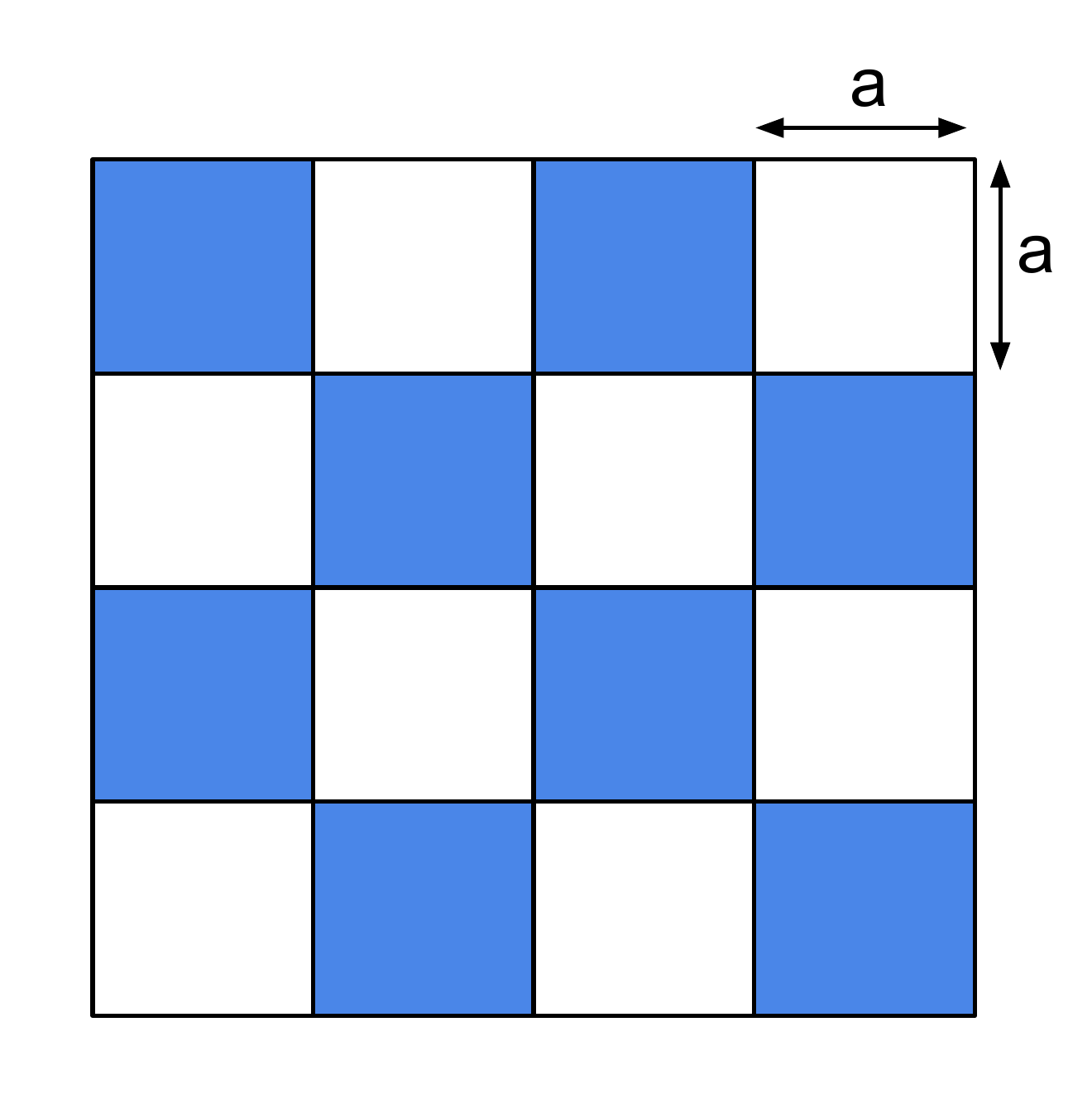}
\caption{(color online) The checkerboard partition with each pixel of side length $a$ is employed in the paper to study the ``band" structure of bulk entanglement spectrum.}\label{p2}
\end{figure}


The approach of solving the integral equation such as Eq.~\ref{integral_eq} becomes much more difficult when the partition is not the left-right one. Now we shall demonstrate that Eq.~\ref{integral_eq} is equivalent to the following eigenvalue equation of projection operator $P$, 

\be
	P|\alpha\rangle = \alpha|\alpha\rangle\:,\label{projector_eq}
\ee along with the particular definition of inner product of states in Hilbert space, 

\be
	\langle \langle a|a' \rangle \rangle := \int_{{\bf r}\in A} d{\bf r}f_{a}^*({\bf r}) f_{a'}({\bf r})\:,\label{inner_product}
\ee associated with any pair of eigenstates $|a\rangle$ and $|a'\rangle$ of physical Hamiltonian. The above definition of inner product differs from the original one in that the range is restricted to the subsystem $A$ of partition. Without the restriction, the eigenstates of physical Hamiltonian are also eigenstates of $P$ with eigenvalue being either $1$ or $0$ depending on wether it is occupied or not.~\cite{Fukui} However, it is the restriction that makes the complete set $\{|a\rangle\}$ {\it not} an orthonormal one, and hence the eigenvalues are generally $0\leq\alpha\leq1$.



Now we introduce the notion of the guiding center representation for correlation matrix. Assuming the size of the two-dimensional space is $L_x\times L_y$, the Landau level states are represented by the ket $|n,X\rangle$ with energy index $n=0,1,2,...$ and $X$ being guiding center coordinate. If the Landau gauge is used, the wavefunction is $\Psi_{nX}({\bf r})=(1/L_y)^{1/2}\exp(iXy/\ell^2)\phi_n(x-X)$ where $\phi_n(x)$ is a one-dimensional harmonic-oscillator eigenstate and $\ell=(\hbar c/eH)^{1/2}$ is the magnetic length. The projection operator associated with filling factor $\nu$ reads,

\be
	P_{\nu} = \sum_{n=0}^{\nu-1}\sum_X |nX\rangle \langle nX|\:.
\ee First consider $\nu=1$ and the left-right partition. It is clear that $\langle \langle0,X'|0,X \rangle\rangle\propto\delta_{X,X'}$ as a result of the preserved translational symmetry along the boundary of partition which is chosen to coincide with the y direction in the Landau gauge. Thus, applying $P_1$ to $|0,X\rangle$ gives the eigenvalue in terms of error function $\alpha_X=\langle\langle 0,X|0,X\rangle\rangle=[1+{\rm Erf}(X/\ell^2)]/2$, which is positive and less than unity. In particular, $\alpha_0=1/2$ shall, according to Eq.~\ref{Cheong1}, correspond to a zero in entanglement spectrum. Hence the approach of Eq.~\ref{projector_eq} is a valid one.


For a more general partition, the eigenvalue equation Eq.~\ref{integral_eq}, following from Eq.~\ref{projector_eq},  becomes 

\be
	\sum_{n',X'} C_{n,X;n',X'}\; s^{(\alpha)}_{n',X'}=\alpha \; s^{(\alpha)}_{n,X} \:,
\ee in which $s$ denotes eigenvector in the guiding center representation, i.e., $s^{(\alpha)}_{n,X}:=\langle \langle n,X|\alpha\rangle \rangle$, and 
the correlation matrix in this representation is

\be
	C_{n,X;n',X'}:=  \langle \langle n,X|n',X' \rangle \rangle \:.\label{S}
\ee  
Here we have adopted the inner product defined in Eq.~\ref{inner_product} so that $C$ is {\it not} the identity matrix. 
Consequently, $\alpha$ is simply the eigenvalue of $C$ in Eq.~\ref{S}, and the corresponding entanglement energy is given by $E_e=\ln\frac{1-\alpha}{\alpha}$.

With the general approach outlined above, we are ready to focus on the entanglement spectrum associated with the checkerboard partition shown in Fig.~\ref{p2}. Unlike the left-right partition in which the matrix $C$ is diagonal with respect to guiding center $X$, the checkerboard partition allows the couplings between different guiding enters $X$ and $X'$ as long as the separation between them is an odd multiple of $G\ell^2$ with $G=\pi/a$ the reciprocal lattice vector. The square pixel has the area of $a^2$. This can be understood from the explicit calculations of $C$, which reads 

\begin{widetext}
\bea
	C_{n,X;n',X'}&=& \int d{\bf r} \left[U_1(x)U_1(y)+U_2(x)U_2(y)\right]
	\Psi^*_{nX}({\bf r})\Psi_{n'X'}({\bf r})\nonumber\\
	&=& \frac{1}{2}\delta_{n,n'}\delta_{X,X'} +i\sum_{m\in\mathbb Z} \frac{\delta_{X',X+(2m-1)G\ell^2}}{(2m-1)\pi}\lambda(n,X;n',X')\:,\label{S1}
\eea in which the restriction to the region in white  (Fig.~\ref{p2}) is replaced by the series of products of Heaviside functions $U_1(x)=\sum_{m\in\mathbb Z}\left\{\Theta(x-2ma)-\Theta[x-(2m+1)a]\right\}$ and $U_2(x)=U_1(x+a)$. The diagonal term of $\frac{1}{2}$ appears due to the fact that all white squares collectively occupy half of the total space and that the probability $|\Psi_{nX}|^2$ is uniform along y direction. The constant $\lambda$ appeared in the second line is a real number, which can be calculated with the following integral,
\bea
	\lambda(n,X;n',X') &=& \int dx\ [U_1(x)-U_2(x)]\phi_n(x-X)\phi_{n'}(x-X')\\
	&=&\frac{n!}{(n')!}\Im\sum_{p=1,3,5,...}\frac{4}{p\pi} \exp\left[ipG\frac{X+X'}{2}\right]F_{n',n}\left[pG,\frac{X'-X}{\ell^2}\right]\:,\label{hopping}
\eea 
\end{widetext} in which the symbol $\Im$ refers to taking the imaginary part of what follows. Thus, the correlation matrix $C$ in guiding center representation resembles a one-dimensional hopping Hamiltonian. We may expand the square wave into the Fourier series, namely $U_1(x)-U_2(x)=\sum_p[4\sin(pGx)]/(p\pi)$ over $p=1,3,5,...$. The overlap of two separate Gaussian functions is often encountered in the context of two-dimensional interacting electrons in the presence of perpendicular magnetic field,~\cite{MacDonald,Luck1} and the function $F_{n',n}(q_x,q_y)=(n!/n'!)[(-q_y+iq_x)\ell/\sqrt{2}]^{n'-n}\exp{(-q^2\ell^2/4)}\mathcal L^{n'-n}_n(q^2\ell^2/2)$ with $\mathcal L$ the associated Laguerre polynomial.

It is of crucial importance to note that matrix $C$ in guiding center representation is Hermitian and that the off diagonal part $C-\frac{1}{2}\mathbb I$ is pure imaginary. Therefore, the entanglement Hamiltonian $H_e'=\ln[(\mathbb I-C)C^{-1}]$ has the particle-hole symmetry under $\mathcal K (\mathbb I-C)\mathcal K=C$ with $\mathcal K$ the complex conjugation.

\section{Entanglement Spectrum}\label{espectrum}

\begin{figure}
\input{epsf}
\includegraphics[width=0.5\textwidth]{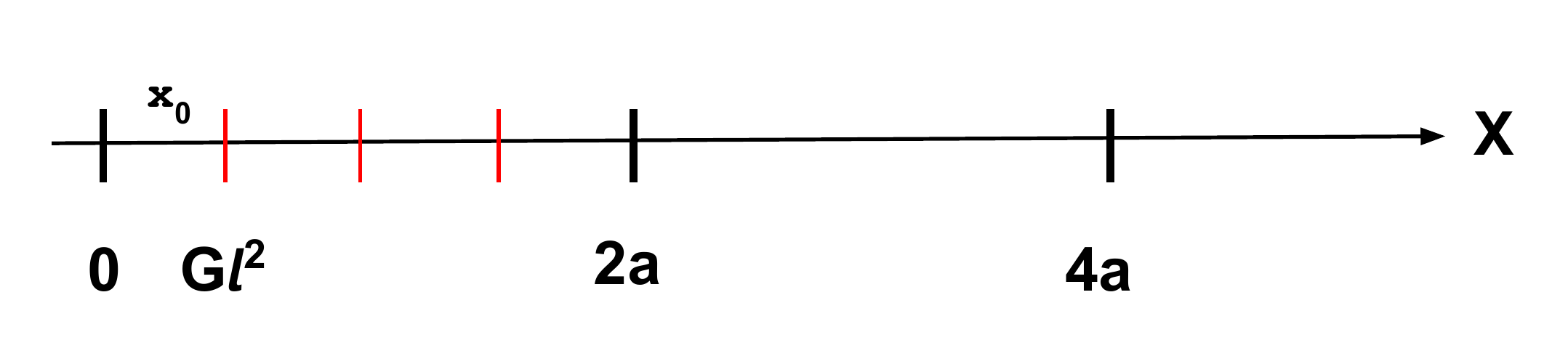}
\caption{The continuous space of guiding center $X$ can be measured by the two length scales $2a$ and $G\ell^2$.}\label{grid}
\end{figure}

Now we are in the position to obtain the entanglement spectrum from the matrix $C$ in Eq.~\ref{S1}. Essentially, $C$ is a huge matrix and can be thought of as motion of hopping in the one-dimensional space of $X$ shown in Fig.~\ref{grid}. Moreover, as a consequence of the checkerboard partition, hopping only occurs between $X$ and $X\pm G\ell^2$, $X\pm 3G\ell^2$, and so on. In addition, it is obvious that $C$ is invariant under the following action of shift by $2a$, 

\be
	\langle \langle n,X|n',X'\rangle \rangle =\langle  \langle n,X+2a|n',X'+2a \rangle\rangle\:,\label{periodicity}
\ee which can be seen from the sine function in Eq.~\ref{hopping} as well. This periodicity simplifies the the problem since we can change to the Bloch basis. Now, as seen in Fig.~\ref{grid}, we may parameterize the guiding center coordinate as 

\be
	X(x_0,i,j)=x_0+2ia+jG\ell^2\:,
\ee with $0<x_0<G\ell^2$ and $i\in\mathbb Z$. The second integer $j\in\mathbb Z_N$ with 

\be
	N:=\frac{2a}{G\ell^2}=\frac{2}{\pi}\Phi\:,\label{relationN}
\ee which is determined by the flux $\Phi=a^2/\ell^2$ threading each pixel in Fig.~\ref{p2}. It is clear that the index $j$ labels the $N$ bands. By Fourier transforming with respect to the $2a$-periodicity, we obtain the following Bloch basis,


\be
	|n,X\rangle\mapsto|n,{\bf k},j\rangle = \sqrt{\frac{2a}{L_x}}\sum_{l\in\mathbb Z} e^{ik_xX} |n,X(k_y\ell^2,l,j)\rangle\:,\label{basis}
\ee which shall acquire a phase factor $\exp{(ik_x2a)}$ upon translating one period along the x axis. The momentum ${\bf k}$ is defined within the first Brillouin zone $[0,G]\times[0,G]$. With this basis labelled by ${\bf k}$, the $C$ matrix is simplified into a $(\nu N)\times(\nu N)$ one, leading to $\nu N$ bands in the BES subsequently. In the followings, we shall specifically calculate the cases for $\nu=1$ and $\nu=2$ and study the corresponding BES.



\subsection{$\nu=1$}


We consider the situation that the flux threading each pixel is $2\pi$, which corresponds to $N=4$ according to Eq.~\ref{relationN}. In this case, we may suppress the energy index $n$ and the matrix elements $\langle\langle {\bf k'},j'|{\bf k},j\rangle\rangle=\delta_{{\bf k,k'}}\bar C_{\bf k}$. The orthogonality follows from the Bloch basis in Eq.~\ref{basis} and the relation in Eq.~\ref{S1}. For convenience, we may rearrange the column vector such that the order of $j$ is $(0,2,1,3)$. It can be shown that the $4\times4$ matrix is




\be
	\bar C_{{\bf k}} = \left(\begin{array}{cccc}
    	\frac{1}{2} & 0 & A & B\\
    	0 & \frac{1}{2} & Q & R\\ 
    	A^*&Q^*&\frac{1}{2}&0\\
    	B^*&R^*&0&\frac{1}{2}
    	\end{array}\right)\:,\label{S_nu0}
\ee and the matrix elements read,

\bea
	A &=& \sum_{l} \langle \langle X(k_y\ell^2,0,0)|X(k_y\ell^2,l,1)\rangle\rangle e^{ik_x(4l+1)G\ell^2},\label{coA-0}\\
	B &=& \sum_{l}\langle \langle X(k_y\ell^2,0,0)|X(k_y\ell^2,l,3)\rangle \rangle e^{ik_x(4l+3)G\ell^2},\\
	Q &=& \sum_{l} \langle \langle X(k_y\ell^2,0,2)|X(k_y\ell^2,l,1)\rangle \rangle e^{ik_x(4l-1)G\ell^2},\\
	R &=& \sum_{l}\langle \langle X(k_y\ell^2,0,2)|X(k_y\ell^2,l,3)\rangle \rangle e^{ik_x(4l+1)G\ell^2}.
\eea Note that the associated phase in above sums is determined by the separations between guiding center coordinates. In obtaining the above relations, we have used Eq.~\ref{periodicity} and the fact that $2a=4G\ell^2$. In addition, the coefficients are not all independent, and it can be shown that $R=-A$ and $Q=-B$ based on the observation that $\langle\langle X(k_y\ell^2,0,j)|X(k_y\ell^2,l,j')\rangle\rangle$ changes sign if $j$ and $(4l+j')$ are simultaneously shifted by 2.

After some algebra, the two independent coefficients for general ${\bf k}$ read

\bea
	A&=&4i\sum_{p,q}\frac{e^{i\pi\frac{q}{2}\tilde k_x}\sin[\frac{p\pi}{2}(\tilde k_y+\frac{q}{2})]}{pq\pi^2}e^{-\frac{\pi}{8}[p^2+q^2]}\:,\label{coA}\\
	B&=&-4i\sum_{p,q}\frac{e^{-i\pi\frac{q}{2}\tilde k_x}\sin[\frac{p\pi}{2}(\tilde k_y-\frac{q}{2})]}{pq\pi^2}e^{-\frac{\pi}{8}[p^2+q^2]}\:,\label{coB}
\eea in which both sums are over $p=1,3,5,...$ and $q=1,-3,5,-7,...$. Note that the rescaled momentum ${\tilde {\bf k}}:= {\bf k}/G$ is defined. Besides, we may also write 

\be
	\bar C_{\bf k} = \frac{1}{2}\mathbb I_4 + \Im(A)N_1+\Im(B)N_2+\Re(A)M_1+\Re(B)M_2\label{DiracMatrix}
\ee with the representation $N_1=-\sigma_y\otimes\sigma_z$, $N_2=-\sigma_x\otimes\sigma_y$, $M_1=\sigma_x\otimes\sigma_z$, and $M_2=-\sigma_y\otimes\sigma_y$. In order to write down the low-lying part of BES analytically, we only retain the largest terms, which correspond to $p=q=1$, in $A$ and $B$. Eigenvalues of $\bar C_{\bf k}$ are shown to be,

\begin{figure}
\input{epsf}
\includegraphics[width=0.2\textwidth]{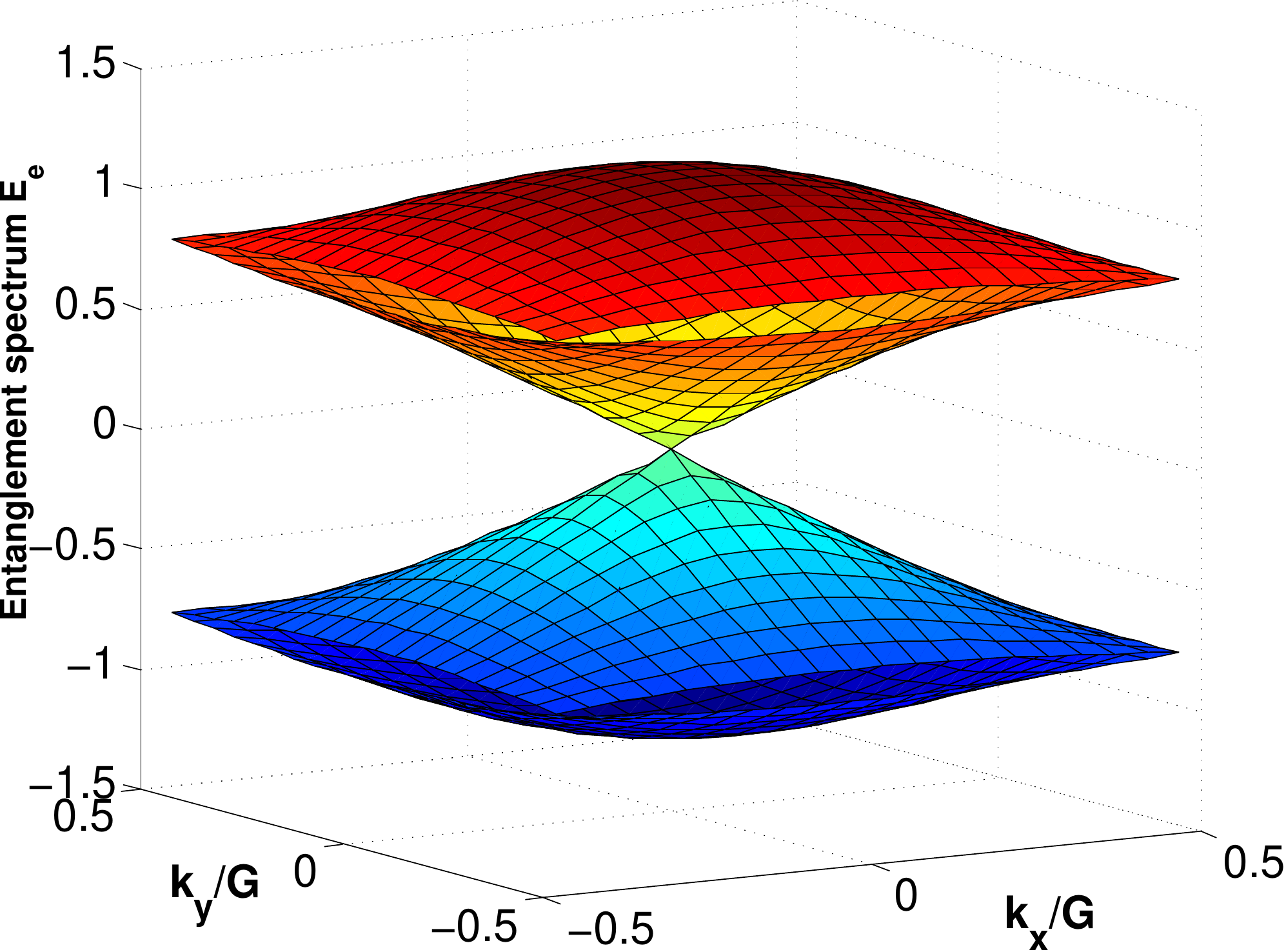}
\includegraphics[width=0.2\textwidth]{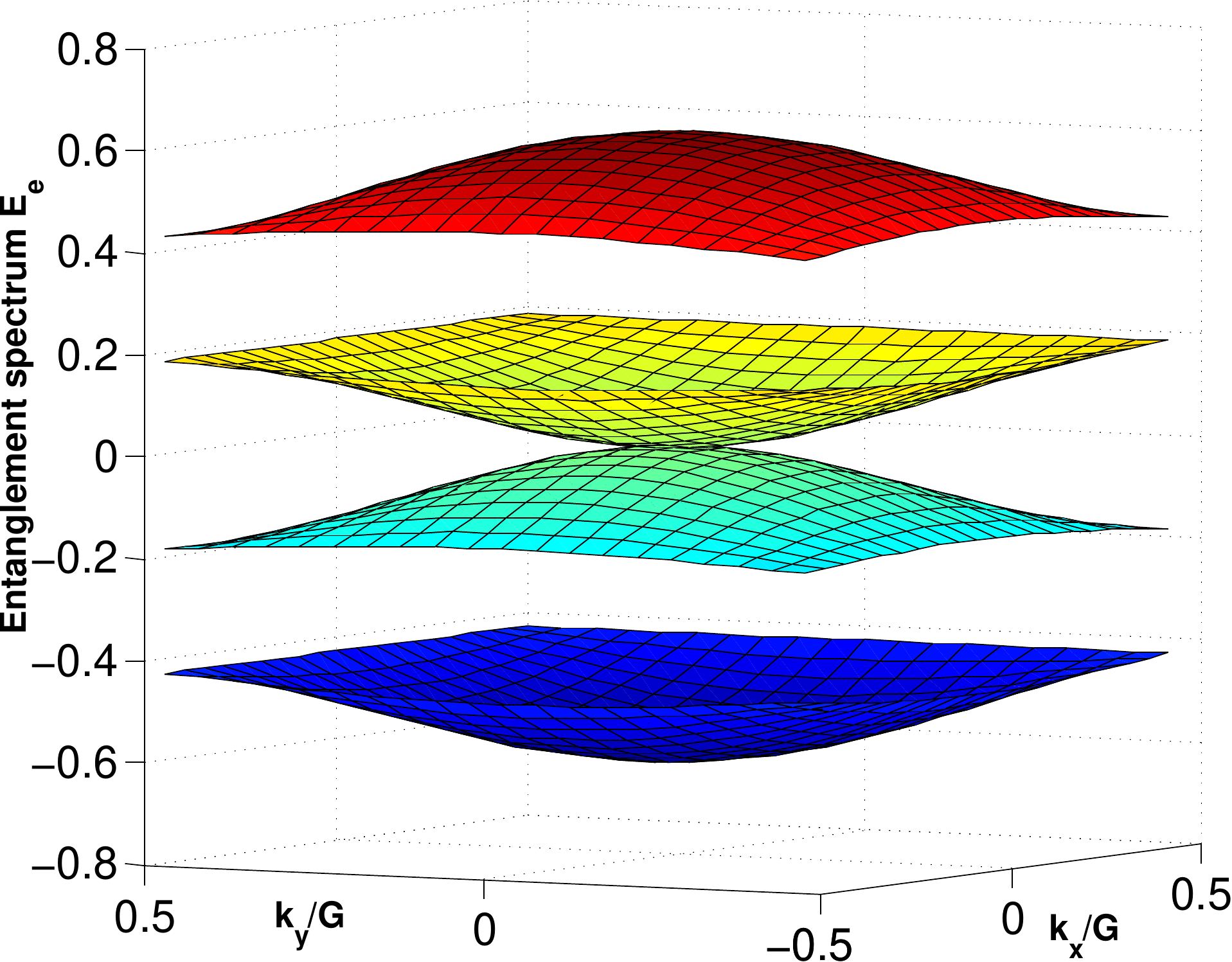}
\caption{(color online) The bands in entanglement spectra $E_e({\bf k})$ of integer quantum Hall state are shown for the parts close to zero energy. Dirac point and quadratic point appear at the origin $\Gamma$ point in the case of $\nu=1$ (left) and $\nu=2$ (right), respectively. 
}\label{iqhe}
\end{figure}

\bea
	\alpha_{\bf k}&\approx& \frac{1}{2} \pm \frac{4}{\pi^2}e^{-\frac{\pi}{4}}\left[
	\sin^2\theta_++\sin^2\theta_-\pm 2\sin\theta_+\sin\theta_-\cos(\pi\tilde k_x)
	\right]^{1/2}\nonumber\\
	&=&\frac{1}{2}\pm\epsilon_{\pm}({\bf k})\:,
\eea  which gives rise to four bands that are symmetric with respect to $\frac{1}{2}$. The angles are $\theta_{\pm}=\frac{\pi}{2}(\tilde k_y\pm\frac{1}{2})$. Then BES is obtained from $E_e({\bf k})=\ln[(1-\alpha_{\bf k})/\alpha_{\bf k}]$. Fig.~\ref{iqhe} displays the numerical results of $E_e(\bf k)$. The Dirac band crossing at zero energy results from taking the plus sign in the bracket, namely $E_e\approx\pm4\epsilon_+({\bf k})\approx\pm8\sqrt{2}\exp(-\pi/4)\tilde k/\pi$.

The momentum ${\bf k}=0$ is a special point at which $A=B$ and $\Re A=\Re B=0$. We may thus write 

\be
	\bar C_{{\bf k}=0}= \frac{1}{2}\mathbb I_4 + \Im(A_0)\left(\begin{array}{cccc}
    	\hat 0 & -\sigma_y(\mathbb I_2+\sigma_x)\\
    	-\sigma_y(\mathbb I_2-\sigma_x)& \hat 0
    	\end{array}\right)\:,
\ee with which we may show that the pair of states, 

\be 
	\varphi_1=\left(\begin{array}{cccc}
    	|{\rm x},+\rangle \\
    	0\\
	0
    	\end{array}\right)\:,\ 
	\varphi_2=\left(\begin{array}{cccc}
    	0\\
	0\\
    	|{\rm x},-\rangle
    	\end{array}\right)\:,\label{zeromode}
\ee are the degenerate eigenstates of $\bar C_{{\bf k}=0}$ with eigenvalue $1/2$. The two-component spinors $|x,\pm\rangle$ are eigenstates of $\sigma_x$ with $\pm$ eigenvalue, respectively. One may immediately see that the two states $\varphi_{1,2}$ correspond to the zero modes of $H_e$ through Eq.~\ref{Cheong1}.


\subsection{$\nu=2$}


Following the above approach, the double-bared ${\bar{\bar C}}_{\bf k}$ denotes the corresponding $8\times8$ matrix for $\nu=2$ integer quantum Hall state in the basis of Eq.~\ref{basis}. It has the form,

\be
	{\bar{\bar C}}_{\bf k} = \left(\begin{array}{cccc}
    	\bar C_{00} & \bar C_{01}\\
    	\bar C_{10} & \bar C_{11}
    	\end{array}\right)\:,\label{C_nu2}
\ee which consists of the $4\times4$ matrices, $\bar C_{00}$ and $\bar C_{11}$, the intra Landau level couplings, and $\bar C_{01}$ and $\bar C_{10}$ ,the inter Landau level couplings. $\bar C_{00}$ is the same as $\bar C_{\bf k}$ in Eq.~\ref{S_nu0}. The difference between $\bar C_{11}$ and $\bar C_{00}$ is that additional factor due to Laguerre polynomial $\mathcal L_1^0$ enters the off-diagonal part of $\bar C_{11}$. It is clear that $\varphi_{1,2}$ are also eigenstates of $\bar C_{11}$. Therefore, in the absence of inter-Landau level couplings $\bar C_{01}$, two Dirac cones shall appear in BES at $\Gamma$ point with the ratio of respective ``Fermi velocity" to be $\eta:=\mathcal L_1^0(G^2\ell^2)=(\pi/2-1)$, approximately. 


Now we are going to show that the BES associated with $\bar{\bar C}_{\bf k}$ has a quadratic band crossing point at ${\bf k}=0$. The situation here is very similar to the bilayer graphene in the sense that the way the two Dirac cones are merged is determined by the ``interlayer" coupling $\bar C_{01}$, which has the general form,

\be
	\bar C_{01} = 
	\left(\begin{array}{cccc}
    	0 & 0 & A' & B'\\
    	0 & 0 & Q' & R'\\
	A'' & B'' & 0 & 0\\
	Q'' & R'' & 0 & 0
    	\end{array}\right)\:.\label{primed}
\ee Unlike $\bar C_{00}$, $\bar C_{01}$ has vanishing diagonal elements because of Eq.~\ref{S1}. Besides, the essence of the presence of a quadratic band crossing point is that at ${\bf k}=0$ all the primed coefficients $A'$, $B'$, $Q'$, and $R'$ vanish while the double-primed ones do not. It could be understood by first inspecting $A'$, which is a sum of $\langle\langle n=0,X(0,0,0)|n=1,X(0,l,1)\rangle\rangle$ over integer $l$. On the other hand, the coefficient $A''$ is a sum of $\langle\langle n=0,X(0,0,1)|n=1,X(0,l,0)\rangle\rangle$. Based on Eq.~\ref{hopping}, it can be shown that at ${\bf k}=0$

\be
	A'(A'')\propto \sum_{p,q} \frac{p\cos pq\frac{\pi}{4}\mp q\sin pq\frac{\pi}{4}}{pq}e^{-\frac{\pi}{8}[p^2+q^2]}\:,
\ee in which the sum is over $p=1,3,5,...$ and $q=1,-3,5,-7,...$ as before. Note that the minus (plus) sign is used exclusively for $A'$ ($A''$). The appearance of both cosine and sine in above expression is due to the fact that the $F$ function in Eq.~\ref{hopping} for inter-Landau level coupling is complex. Now one can easily see that when $p=|q|$ the contributions vanish in $A'$. In addition, the cross terms, for example the pair of contributions from $(p,q)=(1,-3)$ and $(p,q)=(3,1)$, are exactly cancelled in $A'$. In contrast, the coefficient $A''$ is not zero at ${\bf k}=0$. It can be shown that $B'$ vanishes at ${\bf k}=0$ in the same manner. Now we only retain the dominant terms in each coefficient,




\be
	A'=-4i\sin(\frac{\pi}{2}\tilde k_y)\exp(i\frac{\pi}{2}\tilde k_x)\frac{e^{-\frac{\pi}{4}}}{\sqrt{2\pi^3}}=-(B')^*\:,\label{A_prime}
\ee and 

\be
	A''=-4i\cos(\frac{\pi}{2}\tilde k_y)\exp(-i\frac{\pi}{2}\tilde k_x)\frac{e^{-\frac{\pi}{4}}}{\sqrt{2\pi^3}}=-(B'')^*\:.\label{A_dprime}
\ee  Plugging $A''({\bf k}=0)=B''({\bf k}=0)\approx t=-\frac{4i}{\sqrt{2\pi^3}}e^{-\frac{\pi}{4}}$ into ${\bar{\bar C}}_{\bf k}$, the BES of $\nu=2$ state is obtained and is shown in Fig.~\ref{iqhe}. Moreover, one can show with some algebra that the low-lying part of BES is approximately the solution of,

\be
	{\rm det}
	\left(\begin{array}{cccc}
    	\frac{E_e}{4}-\frac{4\epsilon_+^2}{E_e} & 2t \\
	2t^* & \frac{E_e}{4}-\eta^2\frac{4\epsilon_+^2}{E_e}\\
    	\end{array}\right)=0\:,
\ee and it is reminded that $\epsilon_+\propto k$ gives rise to the Dirac point in $\nu=1$ case. Here when $|t|\gg\epsilon_+$ the resultant quadratic band crossing in BES is $E_e\approx\pm2\frac{\eta\epsilon^2_+({\bf k})}{|t|}\propto k^2$.




\subsection{$\pi$-flux}

\begin{figure}
\input{epsf}
\includegraphics[width=0.23\textwidth]{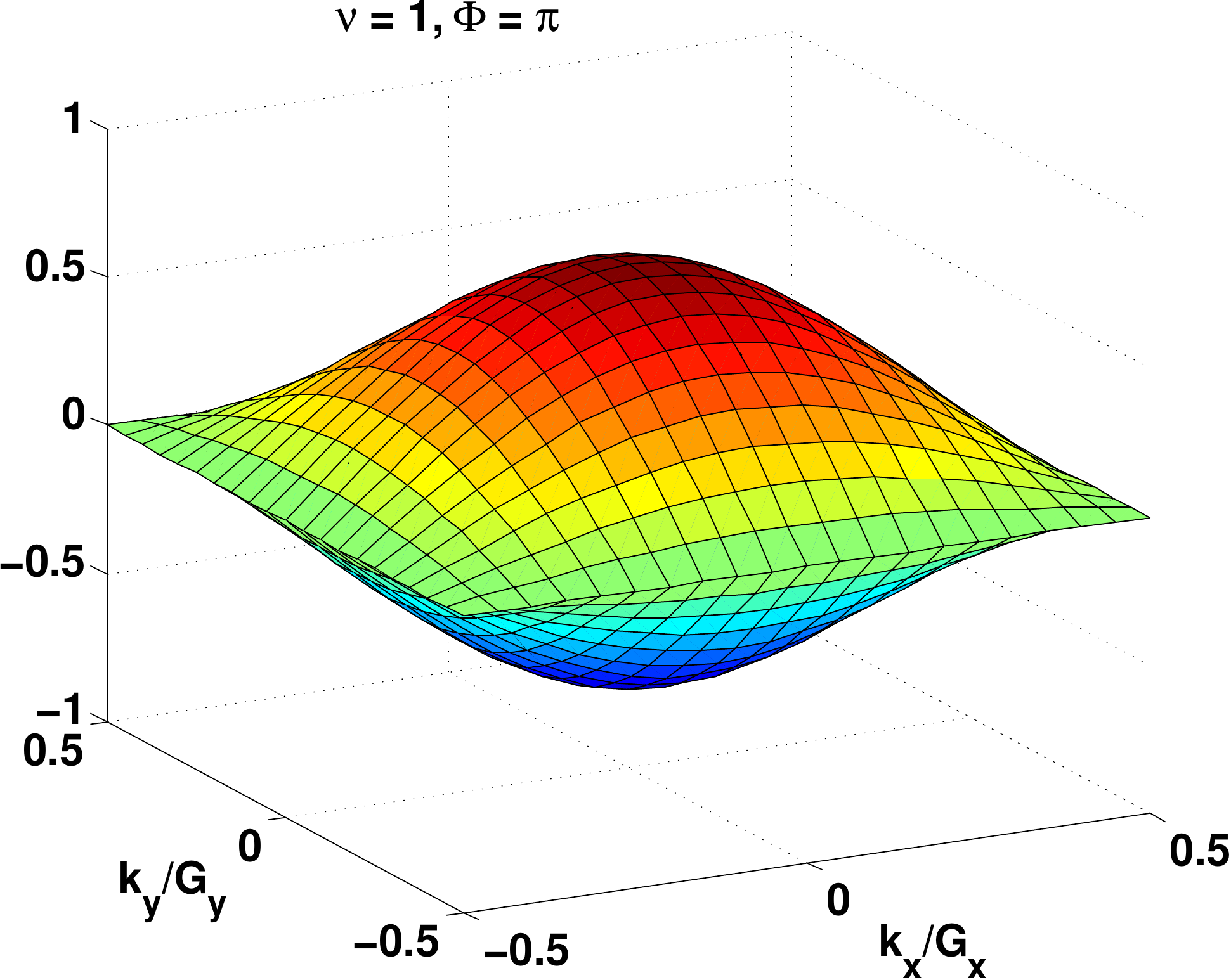}
\includegraphics[width=0.23\textwidth]{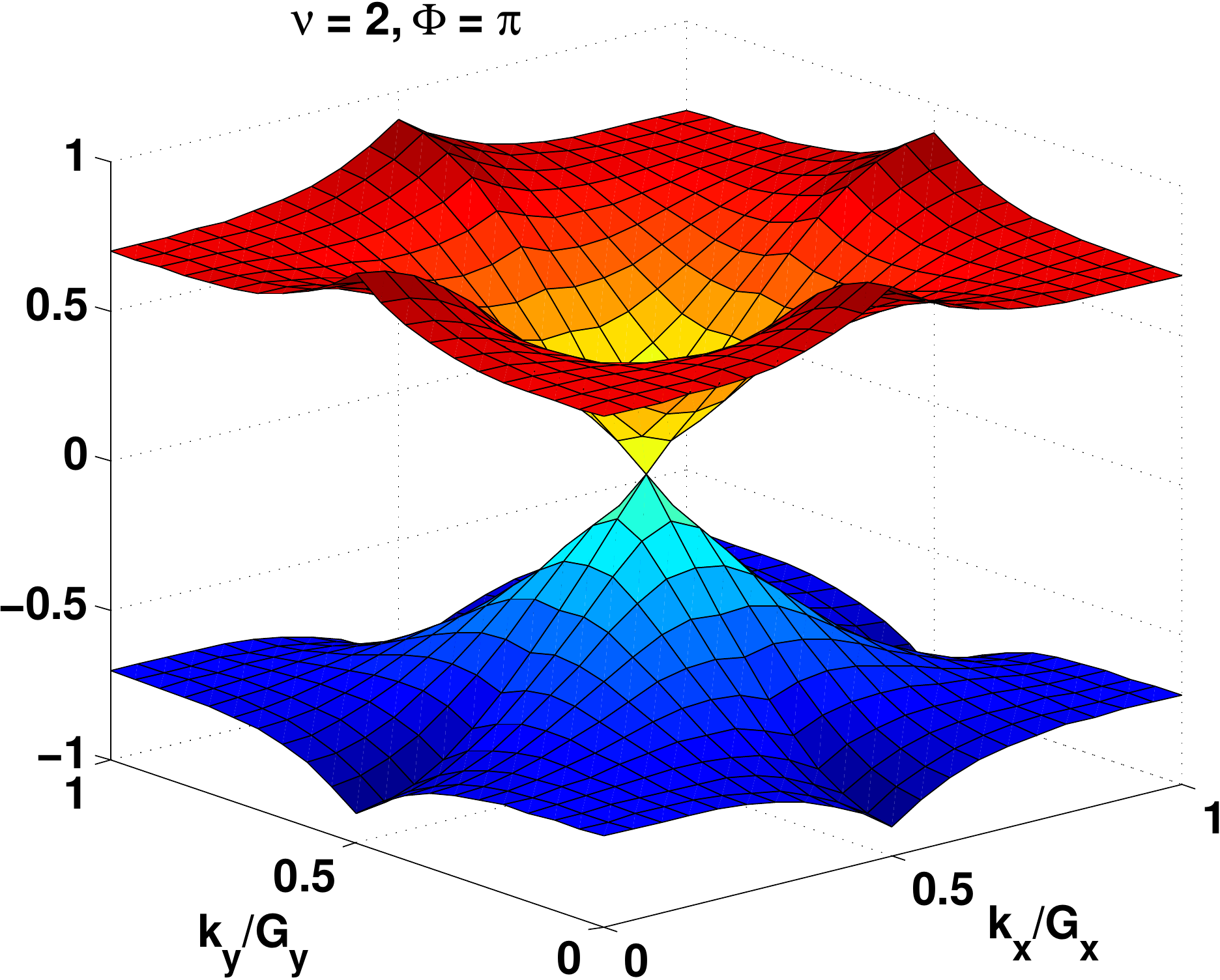}
\caption{(color online) Upper panel: BES of $\nu=1$ and $\Phi=\pi$. Two lines of node appear at $k_x=\pm\frac{G}{2}$ and $k_y=\pm\frac{G}{2}$, and the BES respects the C$_4$ symmetry of checkerboard partition. Lower panel: BES of $\nu=2$ and $\Phi=\pi$. This is equivalent to $\nu=1$ and $\Phi=2\pi$, and the appearance of a single Dirac band crossing located at $(\frac{G}{2},\frac{G}{2})$ confirms the manifestation of Chern number in the band crossing in BES.}\label{flux}
\end{figure}

Here we consider $\Phi=\pi$, which corresponds to $a^2/\ell^2=\pi$ and from Eq.~\ref{relationN} we obtain $N=2$. Thus the corresponding correlation matrix $\tilde C$ is a 2$\times$2 one. For $\nu=1$, it can be shown that $\tilde C=\frac{1}{2}\mathbb I_2+\tilde A\sigma_++{\tilde A}^*\sigma_-$ with 

\be
	\tilde A 
	=  4i\sum_{p,q}\frac{e^{i\pi q\tilde k_x}\sin[p\pi(\tilde k_y+\frac{q}{2})]}{pq\pi^2}e^{-\frac{\pi}{4}[p^2+q^2]}\:,
\ee in which the sum is over $p=1,3,5,...$ and $q=\pm1,\pm3,...$. It can be shown that $\tilde A=0$ along the lines $k_x=\frac{G}{2}$ and $k_y=\frac{G}{2}$. Therefore, as shown in Fig.~\ref{flux}, two lines of node appear in the corresponding BES which respects the C$_4$ symmetry. 

Similar calculations can be carried out for the case of $\nu=2$ and $\Phi=\pi$, which is topologically equivalent to the case of $\nu=1$ and $\Phi=2\pi$. The equivalence is further confirmed by the presence of a single Dirac crossing at ${\bf k}=\frac{G}{2}(1,1)$ shown in Fig.~\ref{flux}.


\subsection{Spectrum symmetries}

As shown in Fig.~\ref{iqhe}, both BES's are symmetric with respect to zero. Here we shall explore the corresponding discrete symmetries in analogy with the time-reversal, chiral, and particle-hole symmetries in the mean-field fermionic Hamiltonian.~\cite{Furusaki,Kitaev} We start with $\nu=1$ case. For $\Phi=2\pi$, let us focus on the matrix $\bar C_{\bf k}$ in Eq.~\ref{DiracMatrix}. First examine the action of $\mathcal K$. Since $\mathcal K (i)\mathcal K = -i$ and $\mathcal K ({\bf k})\mathcal K=-{\bf k}$, one can see that $A$ in Eq.~\ref{coA} and $B$ in Eq.~\ref{coB} are invariant under complex conjugation with ${\bf k}$ defined within BZ. Therefore, the antiunitary operator $I_t=\mathcal K$ commute with $\bar C_{\bf k}$ and we may regard it as the corresponding ``time-reversal symmetry''  in this representation. Moreover, the matrix $\bar C_{\bf k}-\frac{1}{2}\mathbb I$ has only off-diagonal blocks. Therefore, $\Sigma = \sigma_z\otimes\mathbb I_2$ is the ``chiral symmetry'' of BES since $\Sigma^{-1}\bar C_{\bf k}\Sigma=1-{\bar C}_{\bf k}$. Lastly, it is clear that the product $\Xi=\Sigma I_t$ shall serve as the particle-hole symmetry operator. 




For $\nu=2$, the the diagonal blocks, $\bar C_{00}$ and $\bar C_{11}$, of the matrix in Eq.~\ref{C_nu2} share the same spectrum symmetries of $\bar C_{\bf k}$ in Eq.~\ref{DiracMatrix}. So the above three symmetry operators are also the spectrum symmetries of $\bar C_{00}$ and $\bar C_{11}$. Fortunately, the off-diagonal block $\bar C_{01}$ has vanishing diagonal elements, so it also anticommute with $\Sigma$. Consequently, the chiral symmetry is accounted for with the operator $\mathbb I_2\otimes\Sigma$. However, since $I_t$ acts differently on the coefficients $A'$ in Eq.~\ref{A_prime} and $A''$ in Eq.~\ref{A_dprime}, the time-reversal symmetry is broken. 

As for $\nu=1$ and $\Phi=\pi$, it can be seen that the matrix $\tilde C-\frac{1}{2}\mathbb I_2$ commutes with $I_t$ and anticommutes with the chiral symmetry operator $\sigma_z$. The product $\sigma_zI_t$ reflects the BES.

Whether the band crossings in BES are protected can be demonstrated by adding perturbations that break the above symmetries to, for example, Eq.~\ref{DiracMatrix}, following the framework of classifying topological insulators/superconductors.~\cite{Furusaki,Kitaev} However, one has to keep in mind that these additions may not correspond to tunnable parameters in physical system, and only for the purpose of testing the robustness of band crossings. 

\section{Anisotropic partition}

\begin{figure}
\input{epsf}
\includegraphics[width=0.4\textwidth]{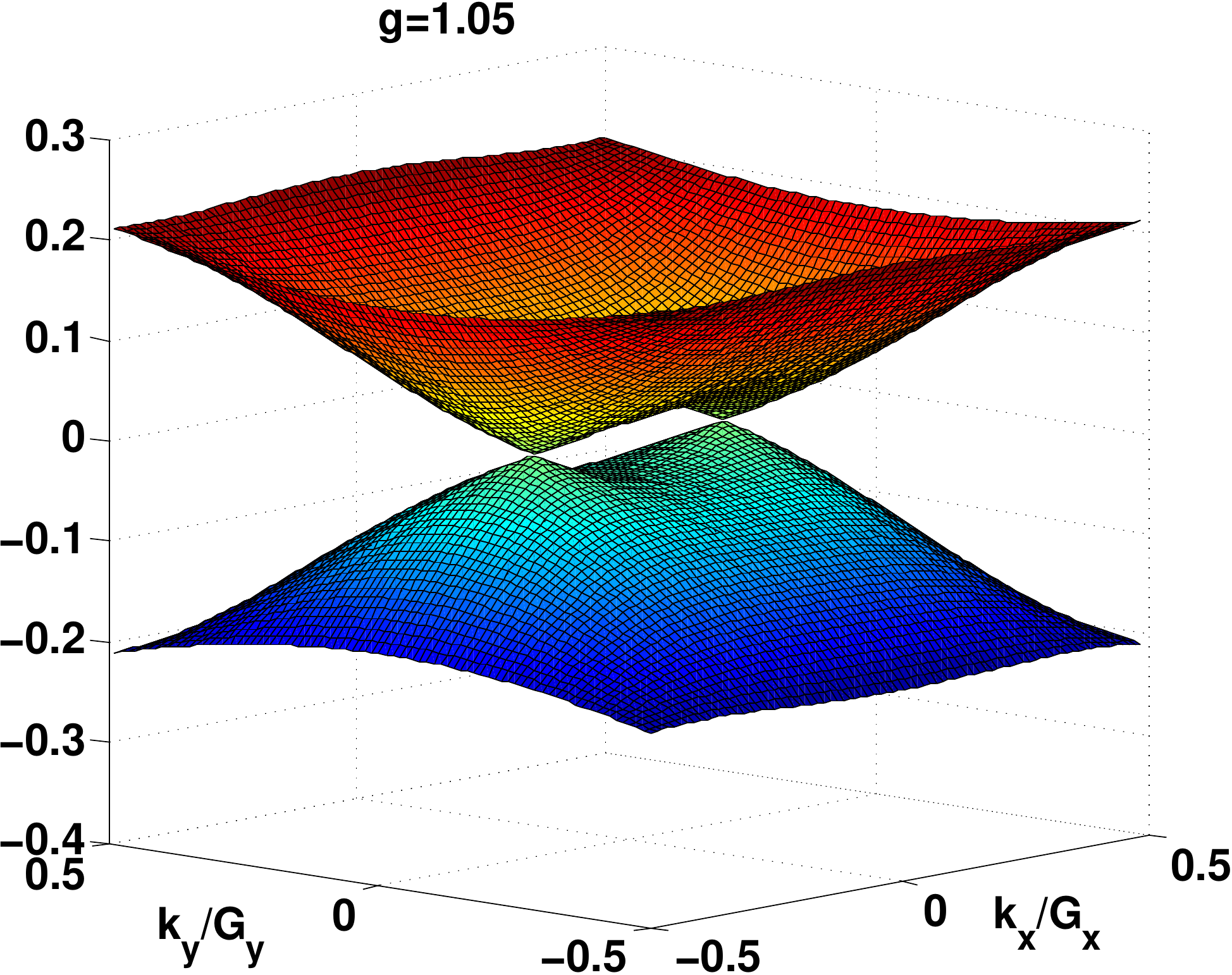}
\caption{(color online) BES of $\nu=2$ and $\Phi=2\pi$ with consideration of anisotropic factor $g:=\frac{b}{a}=1.05$ where $a$ and $b$ are length of sides in x and y directions, respectively. The quadratic band crossing point appeared when $g=1$ is now split into two Dirac points.}\label{g2}
\end{figure}

When the pixel in Fig.~\ref{p2} is of rectangle shape with a ration of two sides $g=\frac{b}{a}>1$, the two reciprocal vectors $G_x=\pi/a$ and $G_y=\pi/b$ are not equal. It follows that $(G_x\ell)^2=g\pi^2/\Phi$ and $(G_y\ell)^2=g^{-1}\pi^2/\Phi$. Thus, the product $G_xG_y\ell^2=\pi/2$ for the flux $\Phi=2\pi$ remains. The breaking of C$_4$ symmetry into a lower C$_2$ symmetry does not ruin the symmetry of intra-Landau level coupling constants. For $\nu=1$, it can be seen from the coefficients in Eq.~\ref{coA} and \ref{coB} that the condition $A=B$ still holds at ${\bf k}=0$ despite the fact that the factor of $\exp{[-\frac{\pi}{8}(p^2+q^2)]}$ in the sum is changed into $\exp{[-\frac{\pi}{8}(gp^2+g^{-1}q^2)]}$. Consequently, the Dirac band crossing in BES is intact, independent of the anisotropic factor $g$. The low-lying spectrum is still isotropic in terms of the rescale momenta $\tilde k_x=k_x/G_x$ and $\tilde k_y=k_y/G_y$.

It is interesting to consider the case of $\nu=2$ and explore the effect of $g$ on the quadratic band crossing. Previously, the vanishing of coupling constants $A'$ and $B'$ at ${\bf k}=0$ is essential for the emergence of quadratic point. However, it is no longer the case when $g\neq1$ since the coupling constants $A'$ and $B'$ are, to the lowest order of $|1-g|$, approximated to be,

\be
	\frac{\pm2i}{\sqrt{\pi^3}} \left[
	g^{\frac{1}{2}}\cos(\frac{\pi}{2}\tilde k_y\pm\frac{\pi}{4})\mp
	g^{-\frac{1}{2}}\sin(\frac{\pi}{2}\tilde k_y\pm\frac{\pi}{4})
	\right]\\
	\times e^{i\frac{\pi}{2}\tilde k_x-\frac{\pi}{4}}\:,
\ee in which the upper (lower) sign is designated for $A'$ ($B'$), respectively. From the numerical result shown in Fig.~\ref{g2}, the factor of $g=1.05$ splits the quadratic point into two Dirac points located along $\tilde k_x$ axis. On further increasing $g$, a larger separation in $\tilde k_x$ occurs. On the other hand, for $g<1$, the splitting then takes place along the $\tilde k_y$ axis.

\bigskip

\section{Conclusions}\label{discussion}

The bulk entanglement spectrum of integer quantum Hall ground states under the checkerboard partition is demonstrated to be a solvable case through our formulation. The manifestation of topological characters, the filling factor $\nu$ and flux $\Phi$ threading each pixel, in band crossings of bulk entanglement spectrum clearly shows the close relation between gapless entanglement spectrum and Chern number of underlying physical system. The appearance of a single Dirac point in either $\nu=1$ and $\Phi=2\pi$ or $\nu=2$ and $\Phi=\pi$ as well as the quadratic point, or two Dirac points, in the case of $\nu=2$ and $\Phi=2\pi$ confirms that the classification in physical system is carried over to the entanglement spectrum. Moreover, the question whether the nodal line enclosing the BZ in the case of $\nu=1$ and $\Phi=\pi$ represents half of a Dirac point is an interesting one. In this paper, we only study the cases with $N=$ 2 and 4 and find the spectra are all symmetric with respect to zero, which is consistent with the fact that the correlation matrix $C$ in Eq.~\ref{S1} can be mapped to $\mathbb I-C$ upon complex conjugation.  On the other hand, it is straightforward to study the cases of $N=1,3,\cdots$, for which our preliminary results show that one band crosses zero but the other $N-1$ bands are gapped but form charge conjugate pairs.

It deserves further study of classifying the topological nature of BES by using the emerging discrete symmetries we have identified, similar to what have been done in classifying the gapless edge modes of topological insulators/superconductors. Moreover, we have only considered the commensurate cases, i.e., $N$ is positive integer, it is also interesting and challenging to solve the BES of the incommensurate cases and find out their physical implication in the similar context of disorder-induced phase transition for the BES of AKLT state studied in the works.~\cite{MIT3,Wan3}

\acknowledgments
Useful correspondences with X. Wan and M.-C. Chung are gratefully appreciated. This work is supported by Taiwan Ministry of Science and Technology through Grant No.~103-2112-M-003-012-MY3, Grant No.~101-2112-M-003-002- MY3 and Grant No.~103-2112-M-003-001-MY3, and also partially supported by National Center for Theoretical Sciences.

\end{document}